\documentclass[sigconf]{acmart}
\renewcommand\footnotetextcopyrightpermission[1]{} 
\setcopyright{none}
\makeatletter
\renewcommand\@formatdoi[1]{\ignorespaces}
\makeatother
\newsavebox\spacewd
\savebox\spacewd{\texttt{ }}
\newenvironment{code}{\par\catcode32=\active \setlength{\parindent}{0pt}\ttfamily}{\par}
{
\catcode32=\active %
\gdef {\makebox[\wd\spacewd][l]{%
\textcolor{white}{\fontfamily{lmtt}\selectfont\large\smash{\char32}}}}%
}
\usepackage{amsmath}
\usepackage{graphicx}
\usepackage{amsfonts}
\usepackage{mathtools}
\usepackage{amsthm}
\usepackage{algorithm,algorithmic}
\usepackage{hyperref}
\usepackage[toc,page]{appendix}

\acmDOI{}
\begin{document}
\pagestyle{plain}
\title{Cyclic Redundancy Checks and Error Detection}

\author{Waylon Jepsen}
\affiliation{
	\institution{Colorado State University}
	\country{waylon.jepsen1@rams.colostate.edu}
}

\begin{abstract}

This study investigates the capabilities of Cyclic Redundancy \\ Checks(CRCs) to detect burst and random errors. Researchers have favored these error detection codes throughout the evolution of computing and have implemented them in communication protocols worldwide. CRCs are integrated into almost every device, in software and hardware. CRCs play a critical role in ensuring that our digital communication systems are efficient and erroneous packets are detected. Because the quantity of data generated and transmitted has increased over the last twenty years, we are more likely to encounter errors. It is important that the tools and methodologies used to ensure the integrity of digital communication systems are evaluated to handle higher frequencies of information.

In this study, we explore the need to improve the capabilities of error-detecting codes to handle higher quantities of data by testing the error detection properties of CRC's in a restricted domain. 
\end{abstract}




\maketitle

\section{Motivation}

As the amount of data communicated over the Internet has increased significantly, the likelihood of experiencing errors has increased. These studies \cite{Feher2011}, \cite{Stone2000}, \cite{Zhuo2017} have suggested that many researchers have either been dealing with corrupted data or are exhausting bandwidth capabilities to overcome from a higher frequency of errors. Undetected errors could be occurring at multiple transmission mediums throughout the network. 

The following work is dedicated to investigating the capabilities of CRCs in the context of higher data rates. High volume networks indicate a greater frequency of error encounters. Interestingly CRCs have remarkable properties when detecting bit errors where the value of a bit changes from a 0 to 1 or vise versa \cite{McNamara1988}. However, errors in the wild can also be bit insertions and deletions \cite{Stone2000} \cite{Zhuo2017} \cite{Feher2011} which introduce challenges to CRCs.

\par Within the scientific community, dealing with corrupt or erroneous data can negatively affect the work of well-intentioned researchers. Data processing in genomics, Machine Learning, and Geo-spatial imaging utilizing large data sets is essential to research. Undetected errors in data could result in faulty research and discrepancies within the scientific community. While the research community is of primary concern, the computing landscape is evolving to handle larger and larger quantities of user data. Thus, the impact upon the commercial computing industry is also of concern. In order to ensure the integrity of large data sets, the error detection mechanisms in place today need to be studied and if appropriate, improved upon. On the other side of the error dilemma, it has become evident that the transmission of packets in large data sets wastes large amounts of high performance bandwidth.  

\section{Prior Work}
There have been few studies on the performance of CRCs in the wild. In 1992 Wang and Crawford studied the capabilities of CRCs in the context of detecting data reordering \cite{wang1992seal}. IEEE CRC-32 was found to be effective at detecting data reordering. In \cite{stone1998performance} CRC and TCP checksums were compared and tested on UNIX file systems and packet sections in asynchronous transfer mode (ATM). The study found that a 10bit-CRC performed as well as a 16 bit TCP checksum and that the TCP checksum detected eliminated cells at a rate of 1 in $2^{10}$ while the CRC-32 performed much better\cite{stone1998performance}. This work shows that CRCs work robustly for data reordering.

In 2002 Koopman evaluated the capabilities of 32-bit CRCs against Hamming distances. Koopman presented the first exhaustive search of 32-bit CRC polynomials and presented domain specific CRC polynomials that provide Hamming distances between 6 and 16k bits and 4 and 114k bits\cite{Koopman2002}. Tridib Chakravarty, one of Koopmans's students, presented an optimized 12bit CRC for short messages (64 bits) for embedded systems that outperform the widely used CCITT 16 bit CRC\cite{chakravarty} under restricted domains. This work shows that for relatively short messages, a proportionally small CRC may perform better.

\subsection{Implementation Optimization}

Recent work has focused on implementation optimization. In 2014 Engdahl and Chung published work on fast parallel implementations for CRC algorithms. Using their software-based parallelism, a speed-up factor of 2.6 was achieved compared to conventional implementations, which involve a table lookup\cite{Engdahl2014}. The proposed improvement processes 32 bits at a time while the traditional table lookup method processes 8-bits at a time. 

The latest work on implementation optimization \cite{Chi2018} presents a table-less algorithm for CRCs. For 16 bit CRCs, the algorithm performs $28\%$ faster than the traditional Sarwate algorithm, which uses a table lookup and is $45\%$ faster for 32 bit CRCs using the Slicing-by-4 algorithm\cite{Chi2018}.

These optimizations increase the speed of CRC implementations, supporting the feasibility of utilizing larger CRC polynomials with a diversity of domain-specific error detection properties. The exploration of CRC polynomials becomes increasing difficult as we explore larger generator polynomials. As the length of the generator increases, the amount of permutations of all possible generator polynomials increases drastically, as exhaustive permutations are $O(n!)$. It has not been proven that the optimization of \cite{Engdahl2014} clearly scales to larger CRCs.

Much of the work on CRC's performance is well over a decade old. Much of the work done in the domain has been theoretical, and few actual experiments evaluating the efficiency of the CRC have been conducted. Prior work has indicated that in actual implementation, the CRC may be underperforming. The following will evaluate the performance of 16 bit CRCs against error patterns classified by Partridge and Stone \cite{Stone2000}.

\section{Introduction}

To detect errors on a transmitted message $m$ of length $n$, a CRC $r$ consisting of $p$ parity bits is appended to the message forming a code $c = [m,r]$ with $k = n+p$  total bits.  The parity bits $r$ are computed such that $r \equiv n (mod (g))$ where $g$ represents the predetermined binary divisor; this is known as the generator. 

The performance of a CRC code depends on its generator. A large body of work in this area is devoted to identifying the qualities of well-performing generators. Let a message $M$ of length $i$ be defined as  $( m_0, m_1,..., m_i )$, defining the corresponding message polynomial as $\sum_{n=0}^{n=i} m_ix^i$. A generator takes on the same form. 

\subsection{Finite Field of Order two}
The mathematical field of this computational domain is the finite field of order two, containing the two elements ${\{0,1}\}$. Finite fields are referred to as Galois Fields and denoted as $GF_n$ where $n$ is the number of elements in the field. $GF_2$ is defined as $\forall n\epsilon \mathbb{N} | n(mod2) \epsilon GF_2 $ where $\mathbb{N}$ is the set of natural numbers. In Computer Science, this field is defined with the XOR operations on binary bits. This allows CRC codes to be implemented efficiently in hardware using a shift register with XOR gates. 

The state-of-the-art is that well-performing generators are primitive. A primitive polynomial generates the extension of all the elements of an extension field from a given base field, where the base field is $GF_2$. Primitive polynomials are irreducible. Additional literature supports the robust mathematical theory regarding the CRC generation process and the classification of generators \cite{Cyclic2007}. This study evaluates a selection of generator polynomials against erasure, insertion, and replacement errors.

\subsection{Examples}

To illustrate the functionality of a CRC, we present a short example. Consider the following information bits 110101101. These bits as the binary coefficients of a polynomial in the finite field of order two are \(x^8+x^7+x^5+x^3+x^2+1\). In order to produce a cyclic redundancy code for this information polynomial, long division is performed with a generator polynomial. For the sake of this example, we will let the generator bits be 10011; thus, the generator polynomial is \(x^4+x+1\). The check appends $n-1$ zero bits to the end of the message, where $n$ is the length of the generator. Long division is performed, giving a remainder bits 0110. The remaining bits are appended to the original message bits so that the generator bits now divide the message bits evenly. The appended bits serve as the checksum which is of length $p-1$ where $p$ is the length of the generator polynomial. The resulting message 11010110\underline{0110} is sent, and the division is performed on the receiving end to check for errors. If the received message does not divide the generator polynomial evenly, the message has an error. 

An illustrative example of the theoretical process is given below. The generator is a four-bit generator $x^3+x^2+1$ corresponding to the bits 1101. The message is 100100. First, $p-1$ zero bits are appended to the message. Then long division is performed under the finite field of order two. 
\begin{code}
         111101 \\
        -----------\\
  1101 | 100100 000 <--- p-1 zero bits\\
         1101|| |||\\
           ----|| |||\\
          1000| |||\\
          1101| |||\\
            ----| |||\\
           1010 |||\\
           1101 |||\\
             ---- |||\\
            111 0||\\
            110 1||\\
              --- -||\\
             01 10|\\
             00 00|\\
              --  --|\\
              1 100\\
              1 101\\
                - ---\\
 Check sum =>   001\\
 append to message => 100100 001 \\
\end{code}
The remainder is appended to the original message serving as the checksum. Notice now that the generator divides the message evenly with no remainder. To illustrate how error detection occurs, assume the message arrives erroneously. The erroneous message is now 110011 001. The recipient of the message then performs the long division check with the generator. Since the Recipient's message will not have a zero remainder, the error is detected.

\subsection{Remediation}
Remediation is achieved depending on the type of media. A \emph{Negative Acknowledgement(NACK)} packet is traditionally sent back to the sender if the receiver's CRC fails \cite{zheng2004multiple} \cite{xmodem}. In some cases re-transmission is carried out. In others, the frame is discarded to allow for continuation of streaming experience. In re-transmission, latency and bandwidth introduce drawbacks that have to be taken into consideration. One such phenomena to consider is the binary exponential back-off algorithm which is used to space out consecutive transmissions of block data to avoid network congestion. Furthermore, the design of the seven layers of the Open Systems Interconnection (OSI) deems that you can't know the remediation protocol at layer $x$ without knowing  the remediation protocol at of layer $x-1$.
\subsection{Properties of CRC check-sums}

CRC codes have a variety of properties regarding the detection of n-bit error patterns. There are two primary classifications of errors that these properties apply to, the first of which is a burst error. The original proposal of CRCs for error detection defines a burst error in the following way.
\begin{definition}
     A \emph{burst error} of length $b$ will be defined as any pattern of errors for which the number of symbols between the first and last errors, including these errors, is $b$  \cite{McNamara1988}.
\end{definition}

For example, a burst error $E(X) = 000\underline{10011}000$ has a length of five. Note that the erroneous bits do not have to start and end with 1. They can also start with one and end with zero. If the zero bit before the first one bit were also erroneous, the error would look like $E(X) = 00\underline{010011}000$. The second classification is a random bit error pattern. 

Additional error classifications are commonly subsets of \emph{Burst} and \emph{Random} errors that support the granularity of errors. It is possible to classify different subsets of burst error patters. Additional classification is impeccably valuable in understanding which errors are native to different data transfer medians.

The burst error detection properties of CRCs are introduced below.

Note that all binary primes other than $2$ begin with a 1-bit and end with a 1-bit. The following are true for primitive generators. If the generator has order length $n$, it detects all burst errors of up to and including length $n$\cite{McNamara1988}.

Because a primitive generator polynomial is irreducible, and the checksum is computed with long division, the only errors that will go undetected are multiples of the generator polynomial. For a message of length $n$ and a CRC of length $k$, the number of times that $k$ divides $n$ is defined as the rate at which undetected errors are expected. The fraction of bursts errors of length $b > n-k$ that are undetectable is $2^{-(n-k)}$\cite{McNamara1988}.

Koopman has shown that polynomials of length $n$ will detect all $(n-1)$ bit errors \cite{koopman}. A generator polynomial with degree $p$ will detect $1-1/2^p$ of all burst errors of length greater than $p+1$ \cite{Cyclic2007} \cite{McNamara1988}. However these properties have not been shown to apply to bit erasure and insertion errors, which have been shown to occur \cite{Stone2000}.

\section{Proposed Approach}

Testing was performed with $26$ 16-bit pre-selected generator polynomials on 727552 erroneous 65536 bit(8192 byte) packets. The errors were fabricated to replicate the errors found in the prior work\cite{Feher2011}, \cite{Stone2000}, \cite{Zhuo2017}. The test was conducted on a LINUX file system in a controlled environment. The machine used is a Dell-R740XD-Xeon-5218 with a 32x2.3G CPU and 768Gb of Memory running Linux(CentOS). The length of the message code being tested was $n=65552$, and the length of our code is $k=16$. Based on prior work, the rate at which we can expect to encounter a single undetected burst error is $1/2^{16}$ or $0.00001525878$, which implies that throughout the study, we can expect to encounter $11.1015625$ undetected errors. The correct CRCs were calculated on the un-corrupted packets and then compared against the erroneous packets. 

\subsection{Polynomial Selection}
When selecting the twenty-seven generator polynomials, efforts were made to select a diverse array of polynomials. This way, the effectiveness of the polynomials could be compared widely. Four different methods were used to select the twenty-seven generator polynomials. 

The first five polynomials were arbitrarily selected from a list of primitive polynomials of length 16 available openly on the web \cite{partow}. These are meant to serve as a control against the other polynomials since the primitive polynomials have performed best in prior work.

In order to select the five consecutive polynomials, primitive polynomials of length 15 were chosen arbitrarily and multiplied by $x+1$ under the binary field. The resulting polynomials served as five generator polynomials of length 16.

The selection of the following ten polynomials was made using a piece of software called Orbiter. Orbiter is a robust open-source mathematical computing software that allows for many operations to be carried out in finite fields. Orbiter is primarily authored by Dr. Betten and is used for the classification of algebraic-geometric and combinatorial objects\cite{abetten}. Orbiter was used to generate ten irreducible polynomials of length 16 over the field $GF_2$.

In collaboration with Allissa Brown, Anton Betten, and Sajeeb Chowdhury, a program was constructed that exhaustively tests polynomials on two-bit error patterns in information polynomials of a specified length. This program was used to test 64,112 16 bit generator polynomials on information polynomials of 64 bits, and five of the best performing polynomials were randomly selected for this study. This selection method will be referred to as AASW for convenience.

In addition, to the previous polynomials, two widely used generator polynomials were selected. The CCITT polynomial used in X.25, V.41, HDLC FCS, XMODEM, Bluetooth, PACTOR, SD, and DigRF was selected for testing. Additionally, IBM's CRC-16 used in Bisync, Modbus, USB, ANSI X3.28, and SIA DC-07 was also selected for testing.

The generator polynomials detailed above were compiled, and the additional information fields concerning the initialization vectors were generated in the table below. This information includes the Polynomials, the binary representation, and the Hexadecimal representation. It is worth noting that one of the 15-bit primitives multiplied by $x+1$ produced the same polynomial as CRC-16 by IBM. This means this polynomial is not primitive. These polynomials were tested using software developed by Dr. Partridge.

\section{Results}


After performing the tests on the selected generator polynomials, the frequency of uncaught errors was compiled to generate the following table. Data analysis can clarify key points. 

\begin{center}
    \begin{table}[ht]
    \centering
        \begin{tabular}{|l|l|l|}
            \hline
            Hex & Uncaught Errors\\ \hline
            0x103DD & 10\\ \hline
            0x1100B & 7\\ \hline
            0x11085 & 14\\ \hline
            0x136C3 & 4\\ \hline
            0x138CB & 12\\ \hline
            0x18005 & 28\\ \hline
            0x18033 & 9\\ \hline
            0x18183 & 16\\ \hline
            0x18151 & 15\\ \hline
            0x18C65 & 13\\ \hline
            0x1322F & 9\\ \hline
            0x1A38D & 15\\ \hline
            0x1B7A9 & 11\\ \hline
            0x14AA7 & 16\\ \hline
            0x1ACD7 & 11\\ \hline
            0x18EB1 & 9\\ \hline
            0x155CF & 10\\ \hline
            0x1B7A5 & 6\\ \hline
            0x170D9 & 9\\ \hline
            0x1D5E3 & 9\\ \hline
            0x11021 & 13\\ \hline
            0x18005 & 28\\ \hline
            0x15FFF & 5\\ \hline
            0x1DFFF & 12\\ \hline
            0x13FFF & 10\\ \hline
            0x1BFFF & 8\\ \hline
            0x17FFF & 8\\ \hline
            
        \end{tabular}
    \end{table}
\end{center}

Plotting the quantity of uncaught errors for each selected polynomial allows us visually digestible information regarding the best and worse performing polynomials. These results suggest that while CRCs are good at detecting burst and bit errors, when erasure and insertion errors occur the error detecting properties of CRCs are reduced. This is intuitively justifiable because the generator polynomial is no longer dividing a bit message string of the expected length.
\begin{center}
    \includegraphics[width=\linewidth]{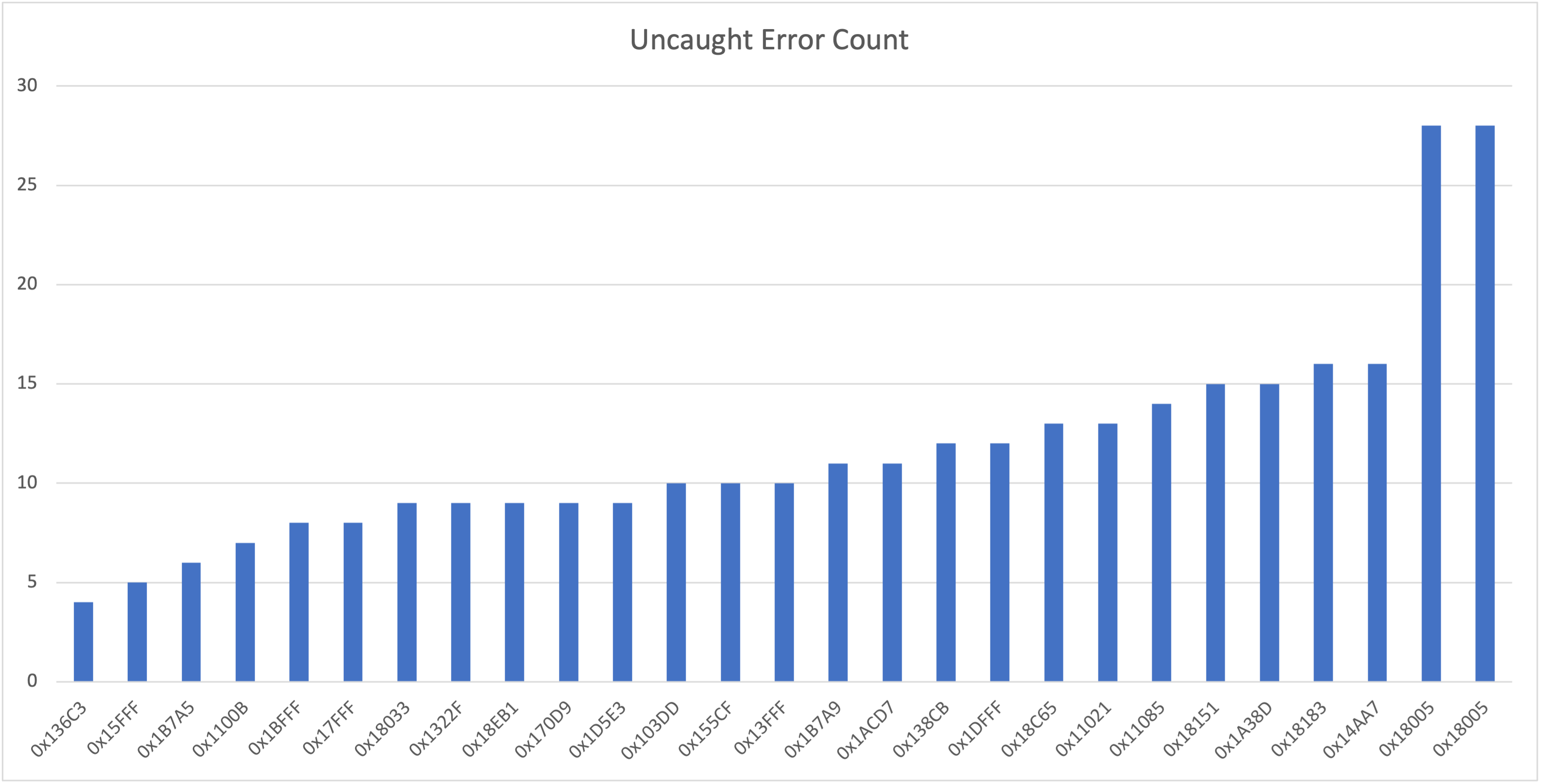}
\end{center}

The study's average number of undetected errors is $11.115385$, which is quite close to our anticipated $11.1015625$. This yields an error of roughly $1.38$\%. (Computed without counting CRC-16 twice). These results validate the capabilities of CRCs of length $n$ missing $1/2^{n}$ of all errors.

\subsection{Best Performing}
As we can see, the two best performing polynomials are
$x^{16}+x^{13}+x^{12}+x^{10}+x^9+x^7+x^6+x^1+1$ and $x^{16}+x^{14}+x^{12}+x^{11}+x^{10}+x^9+x^8+x^7+x^6+x^5+x^4+x^3+x^2+x+1$ Corresponding to 0x136C3 and 0x15FFF. 0x136C3 was an arbitrarily selected primitive polynomial of length 16, so it is expected that this polynomial would perform well. 0x15FFF was selected using the software designed in collaboration with Allissa Brown, Anton Betten, and Sajeeb Chowdhury. It was not hypothesized that these polynomials would perform well.  

\subsection{Worst Performing}
Conversely, the worst-performing polynomials by this metric were $x^{16}+x^{15}+x^2+1$ corresponding to 0x18005 and $x^{16}+x^{14}+x^{11}+x^9+x^7+x^5+x^2+x+1$ corresponding to 0x14AA7. 0x18005 is CRC-16, the widely used CRC by IBM. It is concerning that the CRC-16 polynomial performed poorly, as it is widely implemented and heavily used. 0x14AA7 is one of the irreducible polynomials constructed with Orbiter. 

\subsection{Generator Selection}

An evaluation of four of the primary selection methods was conducted based on the results of the experiment. Seen below is a bar plot of the Methodology against the average number of uncaught errors of the polynomials selected with the corresponding methodology. The worst performing selection method was the selection of primitive polynomials of size $15$ multiplied by $x+1$ over $GF_2$. These results are anticipated as the resulting polynomials would not be irreducible. The best performing selection criteria was AASW described below as "From Program". This was not expected as the AASW testing program does not require the best performing polynomials to be primitive. AASW performed an exhaustive search on polynomials of length 16 and all two bit error patterns in a message polynomial of length 64. You can find the list of generator polynomials used in Table \ref{tab: Generator Polynomials} of Appendix.
\begin{center}
    \includegraphics[scale=.49]{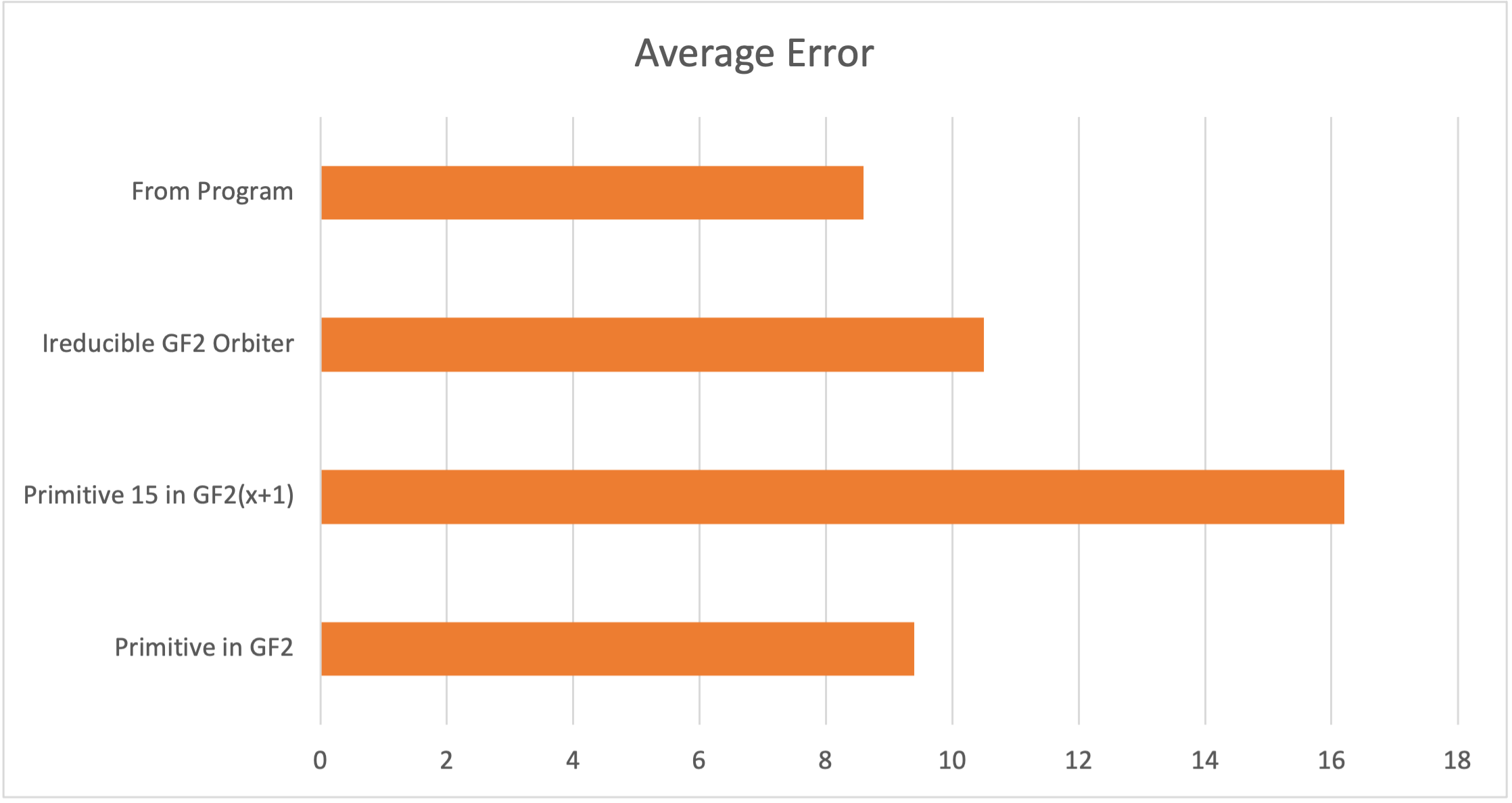}

\end{center}

\subsection{Hamming Distances}

Comparing these results to Koopman's \cite{koopman} we can take a look at the hamming distances of the undetected errors. Koopman has constructed a table available to the public at \cite{koopman} showing the best performing CRCs for specific Hamming distances under the assumption of a low constant random burst error. Koopman's work starts with $HD=2$ and ranges up to $HD=19$ Below is a Box and whisker plot of the hamming distances of the packets that produced undetected errors in the experiment. 
\begin{center}
    \includegraphics[width=\linewidth]{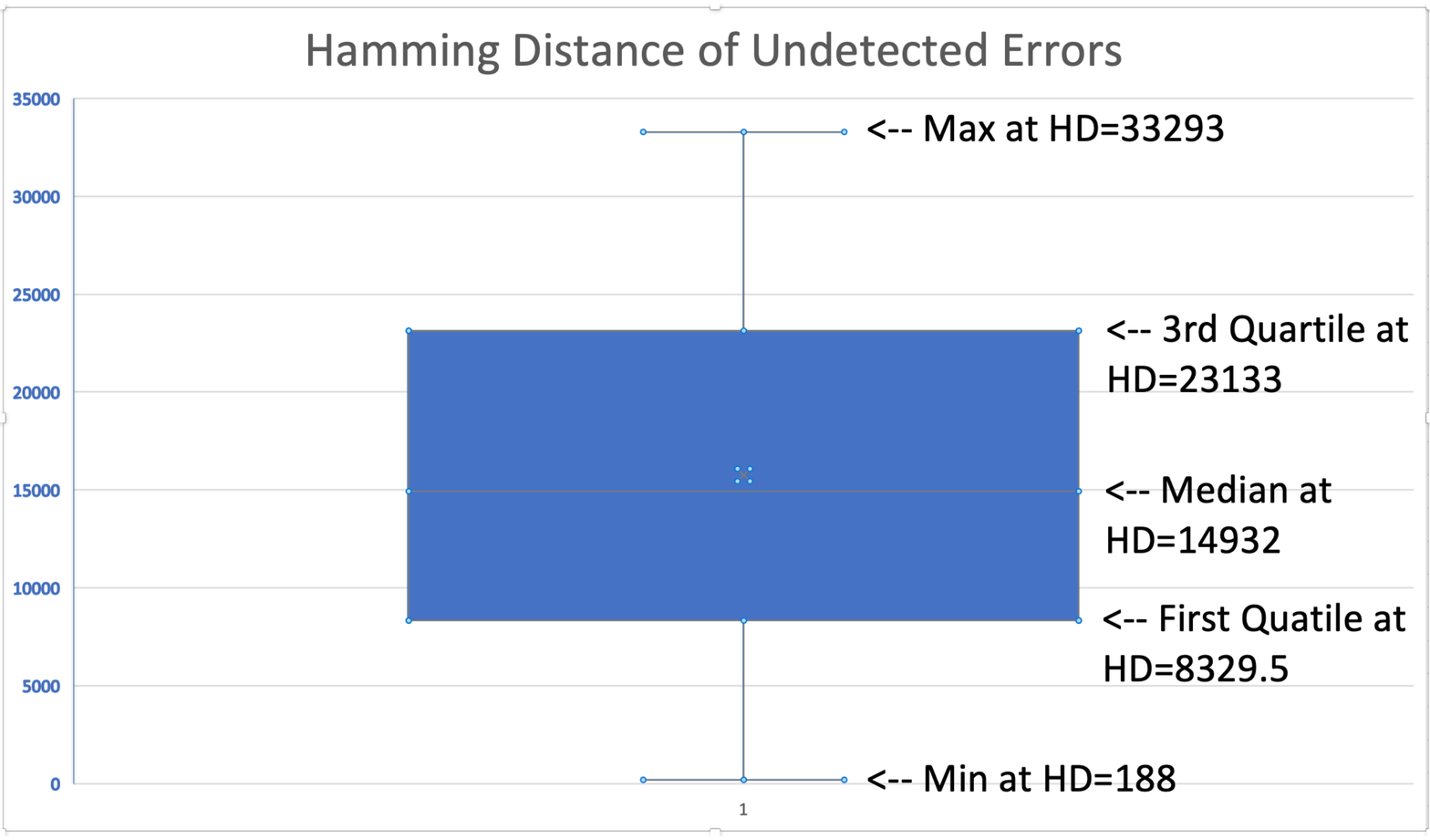}
\end{center}

Here we can see that the smallest undetected error had a hamming distance of $HD=188$, which is well above the scope of Koopman's table. The largest undetected error had a hamming distance of $HD=33293$ while the majority of the undetected errors had hamming distances ranging between $HD=8329.5$ and $HD=23133$. Since these errors were based of errors found in the in \cite{Feher2011}, \cite{Stone2000}, \cite{Zhuo2017}, this supports the claim that erroneous packets in the wild may be more likely to contain larger hamming distances. 

\subsection{Block-156}
Since all of the selected polynomials were tested in the same way, it would be curious if there were any indication of particularly elusive errors. The experiment results have been abstracted for clarity, making it challenging to know all the exact error patterns. However, it was trivial to log the block number of all uncaught errors. Block-156 contained missed errors for five out of the 27 tested generator polynomials. This is more than any of the other blocks, often by a factor of 2. The block has a hamming distance of $7688$ and a total of $1931$ differing bytes. The Hamming distance of this block is below the first quartile of the hamming distance distribution. 

\section{Conclusion}

The CRCs performed as expected in a local LINUX file system. Following the theory of error detection, irreducible polynomials of greater length would increase error detection. Additional data transfer mediums are subject to investigation. Network hardware, such as modems, bridges, and switches, could be culprits worth investigating. Future studies need to be conducted measuring errors over different data mediums to discover indications of justification for uncaught errors.

The average rate of undetected errors for polynomials selected with the AASW program was lower than any other test selection method. Interestingly, the polynomials selected by AASW need not be primitive nor irreducible. Mathematical proofs have been constructed on the error detection benefits of irreducible generators \cite{McNamara1988}\cite{Cyclic2007}. Thus there is an indication that a well-performing generator polynomial may not need to be primitive. Further studies need to be done to verify this claim.

\section{Future Work}
Much is to be done in this field of research. Rigorous evaluations of both statistical computation and mathematical theory are needed to improve our digital communication systems. This study examined a small domain of problems in the greater space of error detection. Replication needs to be carried out accordingly. 

\subsection{AASW Optimization}
Because the AASW program performed the best regarding generator selection, it may be a fruitful endeavor to investigate some optimizations for the program. Since the program runs an exhaustive search, it can take a long time to finish tasks. When the program was run for this experiment with the parameters specified above, the program took an estimated twenty four hours to complete. The program is already optimized to run on GPUs with CUDA, thanks to Sajeeb Chowdhury. However, if strictly being used for test selection, perhaps the program can be optimized further to discontinue the search when a generator fails to catch errors.

\subsection{Tuning the Parameters}
This study has been carried out exclusively with generator polynomials of length 16 against fixed packet sizes. Since the rate depends on the packet size, it is imperative to utilize messages of length resembling standard packet sizes utilized in digital communication systems. With the packet size held at a realistic constant, replicating the study with generator polynomials of the same length and length 32 would be worth investigating. Perhaps it may even be worthwhile to explore the performance of generators of length 64. However, for those aspiring to do so, be prepared to encounter computational restrictions, limiting abilities to do exhaustive searches.

\subsection{Error Classification}

One area of research worthy of further exploration regards the problems of the study of error classification. Traditionally, the research community has classified two types of errors, burst errors and random errors. While these error classifications are undoubtedly helpful and serve to lay the groundwork for mathematical proofs and theorems regarding error detection properties, there is still a problem to be addressed. In order to design efficient error detection systems, higher granularity is needed. Previous work \cite{Stone2000} investigated a variety of error sources and identified 100 error patterns from 500,000 erroneous packets. Partridge and Stone were able to classify about half of the recognized error patterns while the other half remains difficult to classify\cite{Stone2000}. To capture the granular nature of errors in the wild as done in \cite{Stone2000}, gives a starting point domain optimized error detection of increased capabilities. For example, it is hypothesized that the errors relevant in hardware communication systems (USB) may have different properties than errors over TCP/IP. Dr. Partridge has done work evaluating the erroneous packets natural to the TCP/IP/UDP domain, which was utilized in the design of this experiment, but to replicate the same experiment with errors modeled off of hardware transmission could yield different results. 

\subsection{Burst Error Classification Inconsistencies}

Some definitions of burst errors like the one found in \cite{Cyclic2007} have been defined carelessly and maintain inconsistencies with other burst error definitions like those found in \cite{McNamara1988} the original paper proposing CRCs for error detection. In \cite{Cyclic2007} a burst error is defined in the following manner. \emph{Burst of length t} is a vector whose nonzero entries are among t consecutive components, the first and the last of which are nonzero. If
Where as in \cite{McNamara1988} A burst-error of length b will be defined as any pattern of errors for which the number of symbols between the first and last errors, including these errors, is b.
It is recommended that future work stays consistent with \cite{McNamara1988} because it allows for a more accurate representation of what burst errors look like in the wild. In the wild it is not always ensured that the first bit of the burst error will be non-zero. 

\subsection{BCH Codes}

In the presented study, we studied a cyclic checksum which has been standardized. Cyclic codes are a domain in which many more refined codes exist. In coding theory, a Cyclic code is defined as a code in which circular shifts of each codeword result in another code word belonging to that code. A relevant candidate of study is a class of cyclic codes known as BCH codes. BCH codes are constructed using polynomials of a Galois field. A Galois field contains a finite number of elements, for example, the finite field of order two. BCH codes have been utilized in quantum-resistant cryptography\cite{4106108}, satellite communications\cite{891253}, and two dimensional bar codes\cite{4712185}. They have similar advantages to the CRC as they can be easily decoded and implemented in low energy hardware at high speeds. There is worthwhile research to be done in testing BCH codes for error detection on specific error classifications. The hypothesis is that BCH codes may outperform the traditional Cyclic Redundancy Check Sums at detecting a specific classification of errors. This area of exploration has many additional parameters to be explored and optimized regarding the specific nature of BCH codes.

\subsection{Error Correcting Codes}

Error detection is the primary area of study, serving as prerequisite knowledge in designing optimized error correction codes. To anticipate the design of an error correction system without a comprehensive understanding of error detection is a futile endeavor. As the field of error detection evolves, new opportunities arise in developing domain-specific error correction systems. A well-known error correction code is Richard Hamming's Hamming Codes. 

\subsection{Hamming Codes}
A Hamming Code is a class of linear binary codes. Hamming codes are often utilized when error rates are low. They perform well in memory. The length of a binary Hamming codeword is given by $n = 2^r -1$ where r is the number of parity bits. The number of message bits is given by $k = n-r$ \cite{macleod1993coding}. The Hamming codes are implemented in block codes representing bits as a matrix and utilizing parity bits to ensure an even or odd parity in the corresponding rows and columns of the message matrix. A Hamming distance between any two codewords of the same length is commonly defined as the number of indices that differ between the code words\cite{fiedler2004hamming}. This is the metric within which Koopman's study evaluated CRCs \cite{Koopman2002} \cite{chakravarty}. 

\subsection{Example}
For example, the message $10011010$ would be encoded as $\underline{01}1\underline{1}001\underline{0}1010$ where the underlined bits are the parity bits. The position of the parity bit indicates the group of bits it represents. If the parity bit is in position one, it accounts for every other bit in the message. The parity bits assume positions 1, 2, 4, 8.... etc. If the parity bit is 1, then the addition of bits in its group is odd, and 0 if even. The first parity bit is 0 (even). Accounting for every other bit is the code word $P_1$10111 which has an even parity and thus $P_1 = 0$. The second parity bit accounts for 2 bits at a time (check two, skip two, etc), so its group is $P_2$10101, which has an odd parity and thus $P_2 = 0$. The third parity $P_4$ in position 4 counts for all consecutive alternating four bits so it's group is $P_4$0010 and thus $P_4 = 1$. The same methodology applies to the parity bit $P_8$, and if we have a longer message $P_16$ and so forth. 

\subsection{Limitations}
The rate at which Hamming codes can detect errors is given by $R = k/n$, which is equivalent to ${(n-r)}/{(2^r - 1)}$ which is the highest possible for codes with a hamming distance of three. Extended Hamming codes exist, allowing for error detection for 2-bit errors in code words of Hamming distance four and error correction of 1-bit errors. Extended Hamming codes are commonly called Single Error Correcting and Double Error Detecting, often abbreviated as SECDED. Because of their limitations in only detecting minor errors, they are not often utilized in digital communication systems in which significant errors occur. However, they introduce a valuable metric and starting point to build upon.

\section*{Acknowledgment}

I want to acknowledge Dr. Craig Partridge for encouraging this intellectual inquisition and for his careful advising. Additionally, I would like to thank Dr. Anton Betten for his mathematical expertise, and contagious curiosity. I have great appreciation to my peers Alissa Brown, Sajeeb Chowdhury, and Susmit Shannigrahi, who worked together to design one of the elected programs utilized to select generator polynomials for this experiment.

\begin{appendices}
\section{Code}

Providing all the necessary materials to reproduce this experiment, the code the experiment is open source and can be found at \cite{Partridge_CRC_Research_2020}.

\section{Polynomials}
Here we provide the list of generator polynomials with their corresponding binary and hex form below. 

\begin{table*}[t]
    \caption{27 Generator CRCs}
    \label{tab: Generator Polynomials}
    \resizebox{\textwidth}{3in}{%
    \begin{tabular}{|l|l|l|}
    \hline

Polynomial & Binary & Hex\\ \hline
$x^{16}+x^9+x^8+x^7+x^6+x^4+x^3+x^2+1$ & 10000001111011101 & 0x103DD \\ \hline
$x^{16}+x^{12}+x^3+x^1+1$ & 10001000000001011 & 0x1100B \\ \hline
$x^{16}+x^{12}+x^7+x^2+1$ & 10001000010000101 & 0x11085 \\ \hline
$x^{16}+x^{13}+x^{12}+x^{10}+x^9+x^7+x^6+x^1+1$ & 10011011011000011 & 0x136C3 \\ \hline
$x^{16}+x^{13}+x^{12}+x^{11}+x^7+x^6+x^3+x^1+1$ & 10011100011001011 & 0x138CB \\ \hline
$x^{16}+x^{15}+x^2+1$ & 11000000000000101 & 0x18005 \\ \hline
$x^{16}+x^{15}+x^5+x^4+x^1+1$ & 11000000000110011 & 0x18033 \\ \hline
$x^{16}+x^{15}+x^8+x^7+x^1+1$ & 11000000110000011 & 0x18183 \\ \hline
$x^{16}+x^{15}+x^8+x^6+x^4+1$ & 11000000101010001 & 0x18151 \\ \hline
$x^{16}+x^{13}+x^{12}+x^9+x^5+x^3+x^2+x+1$ & 10011001000101111 & 0x1322F \\ \hline
$x^{16}+x^{15}+x^13+x^9+x^8+x^7+x^3+x^2+1$ & 11010001110001101 & 0x1A38D \\ \hline
$x^{16}+x^{15}+x^{13}+x^{12}+x^{10}+x^9+x^8+x^7+x^5+x^3+1$ & 11011011110101001 & 0x1B7A9 \\ \hline
$x^{16}+x^{14}+x^{11}+x^9+x^7+x^5+x^2+x+1$ & 10100101010100111 & 0x14AA7 \\ \hline
$x^{16}+x^{15}+x^{13}+x^{11}+x^{10}+x^7+x^6+x^4+x^2+x+1$ & 11010110011010111 & 0x1ACD7 \\ \hline
$x^{16}+x^{15}+x^{11}+x^{10}+x^9+x^7+x^5+x^4+1$ & 11000111010110001 & 0x18EB1 \\ \hline
$x^{16}+x^{14}+x^{12}+x^{10}+x^8+x^7+x^6+x^3+x^2+x+1$ & 10101010111001111 & 0x155CF \\ \hline
$x^{16}+x^{15}+x^{13}+x^{12}+x^{10}+x^9+x^8+x^7+x^5+x^2+1$ & 11011011110100101 & 0x1B7A5 \\ \hline
$x^{16}+x^{14}+x^{13}+x^{12}+x^7+x^6+x^4+x^3+1$ & 10111000011011001 & 0x170D9 \\ \hline
$x^{16}+x^{15}+x^{14}+x^{12}+x^{10}+x^8+x^7+x^6+x^5+x+1$ & 11101010111100011 & 0x1D5E3 \\ \hline
$x^{16}+x^{12}+x^5+1$ & 10001000000100001 & 0x11021 \\ \hline
$x^{16}+x^{15}+x^2+1$ & 11000000000000101 & 0x18005 \\ \hline
$x^{16}+x^{14}+x^{12}+x^{11}+x^{10}+x^9+x^8+x^7+x^6+x^5+x^4+x^3+x^2+x+1$ & 10101111111111111 & 0x15FFF \\ \hline
$x^{16}+x^{15}+x^{14}+x^{12}+x^{11}+x^{10}+x^9+x^8+x^7+x^6+x^5+x^4+x^3+x^2+x+1$ & 11101111111111111 & 0x1DFFF \\ \hline
$x^16+x^13+x^12+x^11+x^10+x^9+x^8+x^7+x^6+x^5+x^4+x^3+x^2+x+1$ & 10011111111111111 & 0x13FFF \\ \hline
$x^{16}+x^{15}+x^{13}+x^{12}+x^{11}+x^{10}+x^9+x^8+x^7+x^6+x^5+x^4+x^3+x^2+x+1$ & 11011111111111111 & 0x1BFFF \\ \hline
$x^{16}+x^{14}+x^{13}+x^{12}+x^{11}+x^{10}+x^9+x^8+x^7+x^6+x^5+x^4+x^3+x^2+x+1$ & 10111111111111111 & 0x17FFF \\ \hline

    \end{tabular}}
\end{table*}
\clearpage
\end{appendices}

\bibliographystyle{ACM-Reference-Format}
\bibliography{mybib}

\end{document}